\documentclass[aps,prd,onecolumn,showpacs,preprintnumbers,nofootinbib]{revtex4}
\usepackage{amssymb,latexsym}
\usepackage{amsmath,amsbsy}
\usepackage{epsfig,bm}
\DeclareMathOperator{\Pperp}{\mathbf{P}^{\perp}}

\DeclareMathOperator{\kperp}{\mathbf{k}^{\perp}}

\DeclareMathOperator{\dperp}{\mathbf{\Delta}^\perp}
\DeclareMathOperator{\dw}{D_{\text{W}}}
\DeclareMathOperator{\res}{\text{Res}}
\DeclareMathOperator{\kpperp}{\mathbf{k}^{\prime \perp}}
\DeclareMathOperator{\xpp}{x^{\prime \prime}}
\DeclareMathOperator{\xp}{x^\prime}

\DeclareMathOperator{\eps}{\epsilon}
\DeclareMathOperator{\epsp}{\epsilon^\prime}
\DeclareMathOperator{\epspp}{\epsilon^{\prime\prime}}
\DeclareMathOperator{\kppperp}{\mathbf{k}^{\prime \prime \perp}}

\begin{document}
\preprint{NT-UW 03-09}
\title{Complex conjugate poles and parton distributions}
\author{B.~C.~Tiburzi}
\author{W.~Detmold}
\author{G.~A.~Miller}
\affiliation{Department of Physics,  
	University of Washington\\      
	Box 351560
	Seattle, WA 98195-1560}
\date{\today}

\begin{abstract}
We calculate parton and generalized parton distributions in Minkowski space using a scalar propagator with a pair of complex conjugate 
poles. Correct spectral and support properties are obtained only after careful analytic continuation from Euclidean space. Alternately
the quark distribution function can be calculated from modified cutting rules, which put the intermediate state on its complex mass
shells. Distribution functions agree with those resulting from the model's Euclidean space double distribution which we calculate 
via non-diagonal matrix elements of twist-two operators. Thus one can use a wide class of analytic parameterizations 
of the quark propagator to connect Euclidean space Green functions to light-cone dominated amplitudes.  
\end{abstract}

\pacs{13.40.Gp, 13.60.Fz, 14.40.Aq}

\maketitle

\section{Introduction} 
Understanding strong QCD aspects of hadron properties remains a challenge. Experimental probes at large momentum transfer will continue 
to yield 
a wealth of data on hadron structure---stimulating theoretical explanations of the underlying physics in terms of quark and gluon degrees
of freedom. It is well known that light-cone correlation functions are relevant for describing hard processes 
\cite{Lepage:1980fj}, since struck constituents rebound at speeds near that of light. In recent years there has been renewed interest 
in the connection between inclusive and exclusive reactions at large momentum transfer \cite{Muller:1998fv}. The underlying connection
is encompassed by generalized parton distributions (GPDs) which are functions that enter in the description of a variety hard 
exclusive processes \cite{Ji:1998pc}. 

Attempting to describe the non-perturbative light-cone correlations that enter in large momentum transfer processes 
has led to the formulation of gauge theories on the light cone \cite{Brodsky:1997de}. 
There has been progress in directly solving for pion light-cone Fock components 
in light-cone Hamiltonian QCD \cite{Dalley:1998bj}. Another important stride has been made in enumerating and classifying 
hadronic light-cone Fock-space amplitudes \cite{Ji:2002xn}. These could be used for modeling hadrons, although Lorentz covariance 
requires infinitely many Fock components and the incorporation of symmetries into model wave functions has been very limited. 
 
In a different approach, QCD models based on solutions to  Dyson-Schwinger equations provide a useful framework for exploring 
strongly interacting bound states, see e.g.~\cite{Maris:2003vk}.
This framework is fully Poincar\'e covariant, allows for close contact with lattice simulations and provides a means to preserve
symmetries and implement quark and gluon confinement. 
Dyson-Schwinger models have also been used to study light-cone dominated amplitudes. 
Calculation of quark distributions in the impulse approximation 
was undertaken in \cite{Kusaka:1996vm}, where the nucleon Bethe-Salpeter equation was solved in a diquark
spectator model. That investigation relied upon the use of free particle propagators for the quarks and diquark.
In a different study, the authors of Ref.~\cite{Hecht:2000xa} calculated 
quark distribution functions for the pion using a Dyson-Schwinger type model based on entire functions.  Their analysis 
avoided two problems: the integral over the relative light-cone energy is not convergent in the complex plane because 
non-constant entire functions are unbounded; secondly, expressions were derived supposing the 
existence of a K\"all\'en-Lehmann representation which admittedly does not exist for their model propagator.
Although such a model is successful at describing low momentum, space-like processes where the propagator can be approximated 
as entire, the non-analytic points in the whole complex plane must be known to calculate the quark distribution. 
The present investigation enlarges the class of model propagators which can be used to calculate light-cone dominated amplitudes. 
We show how meromorphic propagators can be used in Minkowski space
to arrive at parton and generalized parton distributions as well as the electromagnetic form factor expressed on the light cone. 
Additionally vertex functions for space-like processes can be modeled using meromorphic functions which allows one to go beyond the
impulse approximation. 

Complex conjugate singularities present in solutions to Dyson-Schwinger
equations have been studied in the connection with the violation of Osterwalder-Schrader reflection positivity 
and confinement \cite{Atkinson:1978tk}. 
Recent work \cite{Bhagwat:2003wu} in solving the Bethe-Salpeter equation with a quark propagator consisting of pairs of complex
conjugate singularities shows that the width for meson decay (into free quarks) generated from one pole is exactly canceled by 
the contribution from its complex conjugate. Additionally recent studies have modeled Euclidean space lattice data 
with propagators that have time-like complex conjugate singularities \cite{Bhagwat:2003vw,unp}. In this work, we pursue 
the calculation of space-like amplitudes in Minkowski space using a simple model propagator consisting of a pair of complex 
conjugate poles. Specifically, we are interested in the calculation of light-cone dominated amplitudes for this model which 
necessitates a treatment in Minkowski space.

The paper is organized as follows. First in section \ref{minkowski}, we present the scalar model used to investigate 
complex conjugate poles in propagators. Here we show that naive calculation of the quark distribution in Minkowski 
space is problematic. The distribution has neither proper support nor is positive definite. In section \ref{euclidean} we demonstrate
that despite troubles in Minkowski space, amplitudes involving propagators with pairs of complex conjugate poles can be 
calculated directly in Euclidean 
space. The double distribution (DD) is extracted from non-diagonal matrix elements of twist-two operators. This distribution
satisfies all of the relevant spectral and support properties and can be used to derive parton and generalized parton distributions
for the model, in addition to the electromagnetic form factor. 
In section \ref{cutting}, we show that the quark distribution can be calculated in 
Minkowski space by using a natural modification of the cutting rules applied to the handbag diagram. The effect of these cutting rules
is to put intermediate states on their complex mass shells. Finally in section \ref{analytic}, we rectify the situation in Minkowski 
space by analytically continuing amplitudes from Euclidean space. This justifies the cutting rules presented. The detailed calculation 
of the generalized parton distribution in Minkowski space is contained in the Appendix. After analytic 
continuation, the model's parton and generalized parton distributions agree with those 
calculated from the Euclidean space DD.  A brief summary (section \ref{summy}) concludes the paper.

\section{Problems in Minkowski space} \label{minkowski}
In this section we present the simple model under consideration. Here the propagator is treated in Minkowski space
where difficulties are encountered. We show that amplitudes cannot be directly calculated in Minkowski space when complex conjugate
poles are present in propagators and vertices.

The model we take is $\phi^3$ theory with electromagnetic interactions. Equivalently we can view this model as a bound state of two scalar
particles with a trivial Bethe-Salpeter vertex $\Gamma(k,P) = 1$, where the coupling constant is assumed to be 
absorbed into the overall normalization. We make a simple \emph{Ansatz} for the non-perturbative propagator consisting of a
pair of complex conjugate poles
\begin{equation} \label{Mprop}
S(k)  =  \frac{i (k^2 - a^2 + b^2)}{(k^2 - a^2 + b^2)^2 + 4 a^2 b^2},
\end{equation}
where $a^2 - b^2 > 0$. Defining for ease $m^2 = a^2 - b^2$ and $\epsilon = 2 a b$ (which is taken to be positive without any loss 
of generality), we can write the propagator as
\begin{equation}
S(k) = \frac{i/2}{k^2 - m^2 + i \eps} + \frac{i/2}{k^2 - m^2 - i \eps}.
\end{equation}
 The light-cone energy poles\footnote{For any vector $a^\mu$, we define the light-cone variables $a^\pm \equiv \frac{1}{\sqrt{2}}( a^0 \pm a^3)$.} of the propagator are thus
\begin{equation} \label{apole}
k^-_{a} = \frac{\kperp^2 + m^2}{2 k^+} - \frac{i \eps}{2k^+}
\end{equation}
and $k^-_{a^*} = (k^-_{a})^*$, where $*$ denotes the complex conjugate. Although we use a scalar model, results straightforwardly extend
to spin-$\frac{1}{2}$ particles, e.g., since only the pole structure of Eq.~\eqref{Mprop} is relevant.

\begin{figure}
\begin{center}
\epsfig{file=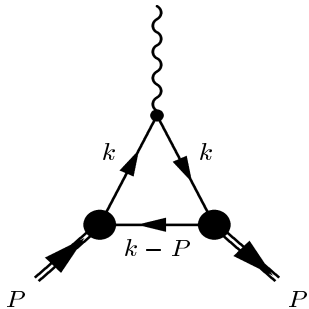,height=1.6in,width=1.6in}
\caption{Triangle diagram at zero momentum transfer used to calculate the quark distribution function by projecting onto the light cone.}
\label{fqofx}
\end{center}
\end{figure}

Now let us consider calculating this model's quark distribution by projecting onto the light cone. The quark distribution
can be derived by fixing the plus-momentum of the active quark $x = k^+/P^+$, see Figure \ref{fqofx}, and taking the plus-component
of the current. Thus up to overall normalization, we have the expression
\begin{equation} \label{minq}
q(x) \propto \int d^4 k \; \delta(k^+ - x P^+) \; x S(k) S(k - P) S(k).
\end{equation}
The $k^-$ integral is then performed by residues. Choosing a frame in which $\Pperp = 0$, the spectator propagator
has light-cone energy poles
\begin{equation} \label{bpole}
k_b^- = P^- + \frac{\kperp^2 + m^2}{2 (k^+ - P^+)} - \frac{i \eps}{2(k^+ - P^+)},
\end{equation}
and $k_{b^*}^- = (k^-_b)^*$. 

Performing the $k^-$ integral in Eq.~\eqref{minq}, we arrive at the quark distribution
\begin{equation}
q(x) \propto 2 \pi  i \Big( -\theta(-x) \big[\res(k^-_{a^*}) + \res(k^-_{b^*})\big] 
+ \theta[x(1-x)] \big[\res(k^-_{a^*}) + \res(k^-_{b}) \big]
+ \theta(x - 1) \big[ \res(k^-_{a^*}) + \res(k^-_{b^*}) \big]  \Big)
\end{equation}
This distribution does not have proper support, i.e.~it is non-vanishing outside the interval $x \in [0,1]$. 
Moreover, the distribution is not real valued, whereas it should be positive definite. Thus the model based 
on the propagator in Eq.~\eqref{Mprop} cannot be suitably formulated in Minkowski space. We will find below that 
Eq.~\eqref{Mprop} makes sense as a Minkowski space propagator only after analytic continuation
from Euclidean space for the amplitude in question.

\section{Covariant calculation in Euclidean space} \label{euclidean}
In Euclidean space, the model propagator is
\begin{equation}
S_{E}(k) = \sum_{\epsilon = \pm } \frac{1/2}{k^2 + m^2 - i \eps}.
\end{equation}
Here and below we use the shorthand $\epsilon = \pm$ to denote the pair of poles $\epsilon = -2 a b, + 2 a b$. 
Unlike in Minkowski space where the measure is imaginary, contributions to Euclidean space amplitudes are real 
and one has no difficulty in calculating form factors 
and distribution functions using $S_{E}(k)$ in the relevant diagrams. The simplicity of the model at hand will allow us to 
calculate its double distribution analytically
and thereby determine the quark distribution and electromagnetic form factor, since these functions are related to the 
double distribution by the so-called reduction relations. The remainder of the paper will 
be devoted to calculation of these quantities in Minkowski space by projecting onto the light-cone.

GPDs are not Lorentz 
invariant objects, however, they stem from a projection of a Lorentz invariant double distribution  function
\cite{Radyushkin:1997ki}. These functions are particularly attractive from the perspective of model building \cite{Mukherjee:2002gb}, 
though one must be careful that the starting point is indeed covariant \cite{Tiburzi:2002kr}, otherwise desirable distribution 
properties and straightforward physical interpretation may be sacrificed. The model under consideration is fully covariant, and thus the 
DD representation is an ideal testing ground for 
our model propagator.  Hence we proceed to calculate the model's Euclidean space DD, recalling along the way the 
relevant properties of DDs.

Let $\overset{\leftrightarrow}{D^\mu} = \overset{\rightarrow}{\partial^\mu} - \overset{\leftarrow}{\partial^\mu} $.
For this scalar model, we define the twist-two operator of spin-$n$ as
\begin{equation}
\mathcal{O}^{\mu \mu_1 \ldots \mu_n} = \phi(0) i \overset{\leftrightarrow}{D} { }^{\big[ \mu}  i\overset{\leftrightarrow}{D}{}^{\mu_1}
\cdots i \overset{\leftrightarrow}{D}{}^{\mu_n}{}^{\big]}    \phi(0),
\end{equation}
where the action of ${}^[ \cdots{}^]$ on Lorentz indices produces only the symmetric traceless part.

We work in Radyushkin's asymmetric frame\footnote{The non-diagonal matrix elements of twist-two operators are, however, 
more conveniently expressed in variables symmetric with respect to initial and final states. Since we use perturbative diagrams
and do not have antiparticles, asymmetrical variables are warranted.
Good discussion of the conversion from symmetrical and asymmetrical 
variables and distributions can be found in \cite{Golec-Biernat:1998ja}. Additionally advantages and disadvantages of both are presented.}
with $P$ as the momentum of the initial state, $P+\Delta$ that of the final and $t = \Delta^2$. Following the two-component formalism of 
\cite{Teryaev:2001qm}, the non-diagonal matrix element of $\mathcal{O}^{\mu \mu_1 \ldots \mu_n}$ can be decomposed into Lorentz invariant moment functions
$A_{n k}(t)$ and $B_{n k}(t)$
\begin{multline} \label{decomp}
\langle \; P + \Delta | \mathcal{O}^{\mu \mu_1 \ldots \mu_n} | P\; \rangle = 
(2 P + \Delta)^{\big[ \mu} \sum_{k = 0}^n \frac{n!}{k!(n-k)!} A_{nk}(t) \; (2P + \Delta)^{\mu_1} \cdots (2P + \Delta)^{\mu_{n-k}} 
(-\Delta)^{\mu_{n - k + 1}} \cdots (-\Delta)^{\mu_n \big]}  \\
- \Delta^{\big[ \mu} \sum_{k = 0}^n \frac{n!}{k!(n-k)!} B_{nk}(t) \; (2 P + \Delta)^{\mu_1} \cdots (2 P + \Delta)^{\mu_{n-k}} 
(-\Delta)^{\mu_{n - k + 1}}  \cdots (-\Delta)^{\mu_n \big]},
\end{multline}
Hermiticity forces the matrix elements of $\mathcal{O}^{\mu \mu_1 \ldots \mu_n}$ to be invariant under the transformation 
\begin{equation} \notag
\begin{cases} 
	P \to P + \Delta \\ 
	\Delta \to - \Delta\\
\end{cases}.  
\end{equation} 
Consequently the values of $k$ are restricted to be even in the first sum  and odd in the second. 
As it stands there is considerable freedom in this decomposition,   e.g.~one could rewrite the above with 
\hbox{$k B_{n,k-1}(t)/ (n - k + 1)$} as  a contribution to $A_{nk}(t)$. 
Carrying this out for all $k$, puts the bulk in the first term and 
renders the second term proportional only to the symmetric traceless part of ($n+1$) $\Delta$'s--- moments of the Polyakov-Weiss 
$D$-term \cite{Polyakov:1999gs}.  This is the usually encountered form of the DD with $D$-term. Alternatively one can also 
express the moments as projections of a single Lorentz invariant function \cite{Belitsky:2000vk}. 
Calculationally, however, Eq.~\eqref{decomp} becomes the most practical to work with \cite{Tiburzi:2002tq}.

The $F$ and $G$ DDs can be defined as generators of the coefficient functions 
\begin{align}
A_{nk}(t) & = \int_{0}^{1} dx \int_{0}^{1 - x} dy \; x^{n - k} (x + 2 y -1 )^k F(x,y;t) \label{Fdef}\\
B_{nk}(t) & = \int_{0}^{1} dx \int_{0}^{1 - x} dy \; x^{n-k}   (x + 2 y - 1)^k G(x,y;t) \label{Gdef}.
\end{align}
As a consequence of the restriction on $k$ in the sums, the function $F(x,y;t)$ is \emph{M\"unchen} 
symmetric \cite{Mankiewicz:1997uy}, i.e.~$F(x,y;t) = F(x,1-x-y;t)$, while $G(x,y;t)$ is \emph{M\"unchen} antisymmetric. 
Also for $n$-even, there is no contribution from the $D$-term to the function $G(x,y;t)$. 

\begin{figure}
\begin{center}
\epsfig{file=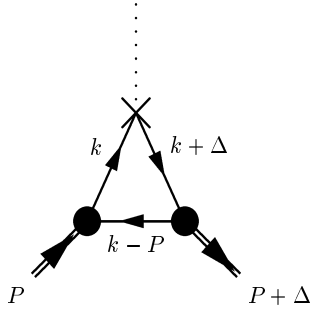,height=1.6in,width=1.6in}
\caption{Diagram used to calculate non-diagonal matrix elements of twist-two operators (denoted by a cross).}
\label{ftwist}
\end{center}
\end{figure}

These functions then appear in the decomposition of  matrix elements of light-like 
separated operators
\begin{multline} \label{DDdecomp}
\langle \; P +\Delta \; | \; \phi(0) \;  i z \cdot \overset{\leftrightarrow}{D}  \; \phi(z^-) \; | \; P \; \rangle \\= 
(2 P \cdot z + \Delta \cdot z) \int_{0}^{1} d x \int_{0}^{1 - x} d y \; 
e^{-i x P \cdot z + i y \Delta \cdot z } F(x,y;t) 
- \Delta \cdot z \int_{0}^{1} d x \int_{0}^{1 - x} d y \; e^{-i x P \cdot z + i y \Delta \cdot z}
G(x,y;t),
\end{multline}
where $z^2 = 0$.

Denoting $\zeta = - \Delta^+/ P^+ > 0$, the GPD in asymmetric variables reads
\begin{equation} \label{GPDdef}
H(x,\zeta,t) =  \int \frac{dz^- e^{i x P^+ z^-}}{2 \pi( 2 - \zeta)} \langle \; P + \Delta  \; | \; \phi(0)  
i \overset{\leftrightarrow}{D^+}  \phi(z^-) \; | \;  P  \; \rangle. 
\end{equation}
Physically $\zeta$ plays the role of Bjorken variable for deeply virtual Compton scattering.
Inserting Eq.~\eqref{DDdecomp} into this definition yields
\begin{equation} \label{JiGPD}
H(x,\zeta,t) = \int_{0}^{1} dz \int_{0}^{1 - z} dy \; \delta(x - z - \zeta y) 
\Big[ F(z,y;t) + \frac{\zeta}{2 - \zeta} G(z,y;t) \Big].  
\end{equation}
By integrating Eq.~\eqref{JiGPD} over $x$, we uncover two sum rules: the sum rule for the form factor
\begin{equation} \label{Fsum}
\int_0^1 d x \int_0^{1-x} d y \; F(x,y;t) = F(t)
\end{equation} 
and the $G$-sum rule
\begin{equation} \label{Gsum}
\int_0^1 d x \int_0^{1-x} d y \; G(x,y;t) = 0,
\end{equation}
which follows since $G$ is \emph{M\"unchen} antisymmetric. Eq.~\eqref{Gsum} is important and mandated by current conservation. 
All too frequently the $G$ DD function is overlooked and treated as identically zero.  
Lastly, the quark distribution function $q(x)$ can be found from the DD at 
zero momentum transfer, see Eq.~\eqref{GPDdef}, 
\begin{equation} \label{qred}
q(x) = \int_0^{1-x} dy F(x,y;0).
\end{equation}

We can use the decomposition in Eq.~\eqref{decomp} 
to calculate our simple model's DD. Parameterizing the momenta as in Figure \ref{ftwist}, the non-diagonal
matrix element of $\mathcal{O}^{(n)}$ reads
\begin{equation} \label{bob}
\langle \; P + \Delta | \mathcal{O}^{\mu \mu_1 \ldots \mu_n} | P\; \rangle  = \frac{2 N}{\pi^2}
\sum_{\eps,\epsp,\epspp = \pm} \int d^4 k \frac{(2 k + \Delta)^{\big[\mu}(2 k + \Delta)^{\mu_1} \cdots (2 k + \Delta)^{\mu_n\big]}}
{[k^2 + m^2 - i \eps] [ (k+\Delta)^2 + m^2 - i \epsp] [(k-P)^2 + m^2 - i \epspp]}
\end{equation}
The normalization constant $N$ is chosen by the condition $F(0) = 1$.
Let us denote the propagators
simply by $\mathfrak{A} = (k - P)^2 + m^2 - i \epspp$, $\mathfrak{B} = (k+\Delta)^2 + m^2 - i \epsp$ and  
$\mathfrak{C} = k^2 + m^2 - i \eps$. We introduce two Feynman parameters $\{x,y\}$ 
to render the denominator specifically in the form $[x \mathfrak{A} + y \mathfrak{B} + (1-x-y) \mathfrak{C}]^{-3}$.
One then translates $k^\mu$ to render the integral (hyper-) spherically symmetric via the definition 
$k^\mu = l^\mu + x P^\mu -  y \Delta^\mu$. The resulting integral over $l$ can be evaluated directly (remember we are in Euclidean space). 

Binomially expanding the result of the integral, we can make contact with Eq.~\eqref{decomp} and subsequently determine the
$F$ and $G$ double distributions by inspection from Eqs.~\eqref{Fdef} and \eqref{Gdef}. Defining the auxiliary functions
\begin{equation} \label{dzero}
D_o(x,y;t) = m^2 - x ( 1- x) M^2 - y (1- x - y) t
\end{equation}
and
\begin{equation}
D(x,y;t) = N \sum_{z = 0, x, y, x+y} \frac{D_o(x,y;t)}{D_o(x,y;t)^2 + \eps^2 (1 - 2 z)^2},
\end{equation}
the DDs can be written simply as
\begin{align}
F(x,y;t) & = x D(x,y;t) \label{FDD} \\
G(x,y;t) & = (x + 2 y - 1) D(x,y;t) \label{GDD}.
\end{align}
Accordingly $F$ is \emph{M\"unchen} symmetric and $G$ is antisymmetric.
Notice although $\epsilon$ is finite, corresponding results using the standard perturbative propagator 
can always be recovered in the limit $\epsilon \to 0$. For example, 
the correct $F$ and $G$ DDs are recovered in the limit $\eps \to 0$ \cite{Tiburzi:2002tq}. 

The model GPD can be derived by utilizing Eq.~\eqref{JiGPD}, although the integral must be performed numerically. The quark distribution
can be found via the reduction relation Eq.~\eqref{qred}, namely
\begin{equation} \label{qofx}
q(x) = N \sum_{z = 0, x}\Bigg( \frac{x(1-x) D_o(x,0;0)}{D_o(x,0;0)^2 + \eps^2 (1 - 2 z)^2}  + \frac{x}{\eps} \tan^{-1} 
 \frac{\eps ( 1 - 2 z) }{D_o(x,0;0)}  \Bigg)
\end{equation}
Lastly the form factor can be found from the sum rule Eq.~\eqref{Fsum}

In Figure $3$, we plot the quark distribution and electromagnetic form factor for various values of $\eps$ in GeV${}^2$. 
We have arbitrarily chosen the other model parameters as $M = 0.14$ GeV and $m = 0.33$ GeV. Additionally in Figure $4$, the
GPD is plotted: first at fixed $\zeta$ and $t$ for various values of $\eps$ and then at fixed $t$ and $\eps$ for various values 
of $\zeta$. Curves corresponding to $\eps = 0$ are the standard results for a propagator with one real pole.

\begin{figure}
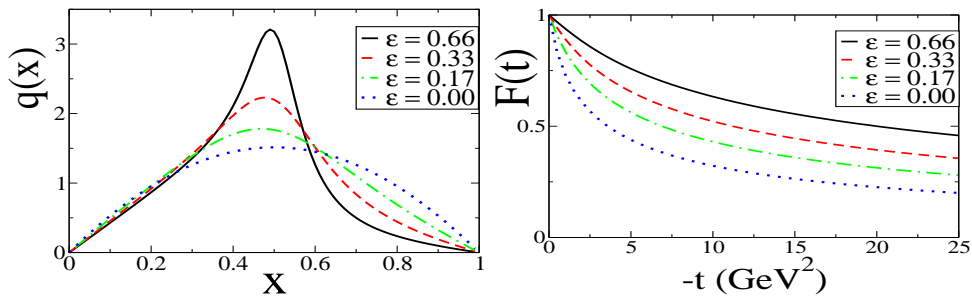

\begin{center}
\epsfig{file=q.eps,height=1.5in,width=2.5in}
\epsfig{file=form.eps,height=1.5in,width=2.5in}
\caption{On the left, the quark distribution Eq.~\eqref{qofx} is plotted as a 
function of $x$ for a few values of $\eps$ in GeV${}^2$.  On the right, the form factor 
calculated from Eqs.~\eqref{Fsum} and \eqref{FDD} is plotted as a function 
of $-t$ for a few values of $\eps$ in GeV${}^2$. 
The model parameters are arbitrarily chosen as: $M = 0.14$ GeV 
and $m = 0.33$ GeV.}
\end{center}
\label{figqandF}
\end{figure}

\begin{figure}
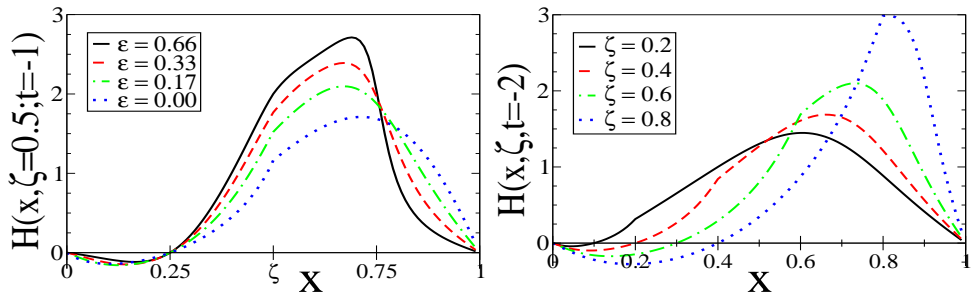

\begin{center}
\epsfig{file=gpd.eps,height=1.5in,width=2.5in}
\epsfig{file=gpd2.eps,height=1.5in,width=2.5in}
\caption{Plots of the GPD calculated from Eqs.~\eqref{FDD},\eqref{GDD} and \eqref{JiGPD}. 
On the left, the GPD is plotted as a 
function of $x$ for a few values of $\eps$ (in Gev${}^2$) at fixed $\zeta = 0.5$ and $t = - 1.0$ GeV${}^2$. On the right, 
the GPD appears at fixed $\eps = 0.17$ GeV${}^2$ and $t = - 2.0$ GeV${}^2$ and is plotted
as a function of $x$ for a few values of $\zeta$.
The model parameters are arbitrarily chosen as: $M = 0.14$ GeV and 
$m = 0.33$ GeV.}
\end{center}
\label{figGPD}
\end{figure}

\section{Cutting rules} \label{cutting}
In this section, we show there is some hope in working with the model propagator Eq.~\eqref{Mprop} in Minkowski space. 
We demonstrate that the quark distribution can be derived by a straightforward generalization of the cutting rules. 

\begin{figure}
\begin{center}
\epsfig{file=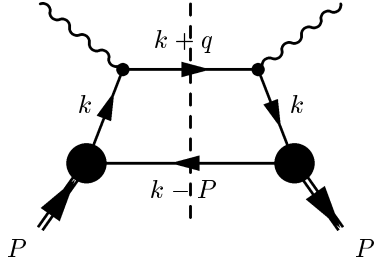}
\caption{Handbag diagram for the forward Compton amplitude. 
Dashed line denotes the cut which yields the quark distribution in the Bjorken limit.}
\label{fhand}
\end{center}
\end{figure}

Consider the forward Compton amplitude $\mathcal{M}^{\mu \nu}$ depicted in Figure \ref{fhand}. In the Bjorken limit, 
the imaginary part of this diagram is related to the quark distribution. For simplicity, we can choose a frame in which 
$\mathbf{q}^\perp = 0$. The minus-plus component of the forward Compton amplitude 
in such frames reads
\begin{equation} \label{fca}
i \mathcal{M}^{-+} = - \frac{16 N}{\pi^3} \int d^4 k \; (2 k^- + q^-)  S(k) \; S(k-P) \; S(k) \; (2 k^+ + q^+) S(k+q). 
\end{equation}

In the scalar particle case, the minus-plus component of the forward Compton amplitude can be used to define the quark distribution 
in a way analogous 
to the spin-$\frac{1}{2}$ case. The relation is simply $\Im (\mathcal{M}^{-+}) \propto q(x)$, see \cite{Tiburzi:2001je}. 
In Eq.~\eqref{fca}, we have adjusted the overall normalization so that equality between $\Im (\mathcal{M}^{-+})$ and $q(x)$ holds. 
With standard perturbative propagators, we could calculate the imaginary part of $\mathcal{M}^{\mu \nu}$
by using the cutting rules, whereby one replaces the cut propagators in Figure \ref{fhand} by an on-shell prescription, namely
\begin{equation} \label{cutrule}
S(k)_{\eps = 0}  \longrightarrow  - 2 \pi i \; \delta (k^2 - m^2).
\end{equation}
In Eq.~\eqref{cutrule}, we have specified $\eps = 0$ for the case of the free particle propagator. Since large momentum flows through
the handle of the handbag, we may neglect the mass of the struck quark and use the standard cutting rule for $S(k+q)$. In the Bjorken
limit, we define $x = - q^2 / 2 P \cdot q$  which remains finite as $q^2, P \cdot q \to \infty$ and is kinematically bounded
between zero and one. Further we orient our frame of reference so that  $q$ has a 
large minus component in this limit, and consequently  $q^+ = q^2 / 2 q^-$ is finite. Hence we have the familiar replacement
\begin{equation} \label{xcut}
S(k+q) \longrightarrow - \frac{\pi i}{2 q^-} \; \delta( k^+ - x P^+). 
\end{equation} 

To complete the cut, we must deal with the spectator particle's complex mass shells. We must worry about the propagator Eq.~\eqref{Mprop}
only where the denominator is zero. 
Thus we are lead to the cutting rule for the propagator Eq.~\eqref{Mprop}
\begin{equation} \label{gencut}
S(k) \longrightarrow - \pi i \Big[  \delta(k^2 - m^2 + i \eps) + \delta(k^2 - m^2 -  i \eps )  \Big],
\end{equation}
which puts the intermediate state on its complex mass shells. Furthermore, the limit $\eps \to 0$ produces
the regular cutting rule Eq.~\eqref{cutrule}. 

Using this cutting rule for the spectator particle along with Eq.~\eqref{xcut}, we can deduce the quark distribution 
from $\Im (\mathcal{M}^{-+})$ in the Bjorken limit
\begin{equation} \label{qcut}
q(x) = \frac{4 N}{\pi} \int d^4 k  \; \delta(k^+ - x P^+) 
\Big[\delta(k^- - k_b^-) + \delta( k^-  - k^-_{b^*}) \Big] \frac{x S(k)^2}{1-x },
\end{equation}
where the light-cone energy poles are given in Eq.~\eqref{bpole}. Notice the resulting distribution is real and has proper
support due to the kinematic constraint $x \in [0,1]$ imposed by the Bjorken limit. Evaluation of the two trivial integrals
leaves only the transverse momentum integration
\begin{equation} \label{LCqx}
q(x) =  \frac{N}{\pi} \sum_{\eps,\epsp,\epspp = \pm} 
\int \frac{d\kperp}{x(1-x)} \dw(x,\kperp, \eps, \epspp |M^2) \dw(x,\kperp,\epsp,\epspp|M^2)
\end{equation}
where we have defined the Weinberg propagator generalized for complex masses as
\begin{equation} \label{weinprop}
\dw(x,\kperp,\eps,\epsp|M^2)^{-1} = M^2 - \frac{\kperp^2 + m^2}{x(1-x)} + \frac{i \eps}{x} + \frac{i \epsp}{1 - x}.
\end{equation}

Evaluation of the $\kperp$ integral yields an analytic
expression for $q(x)$, which is identical to that obtained from the DD Eq.~\eqref{qofx}. Notice Eq.~\eqref{qcut} is equivalent
to evaluating Eq.~\eqref{minq} at $\res(k^-_b) + \res(k^-_{b^*})$ and hence the quark distribution is derived as if the
complex conjugate spectator poles both lie in the upper-half complex plane! We will understand this better once we analytically 
continue from Euclidean space.

\section{Analytic continuation} \label{analytic}
Above we have seen that modified cutting rules can be used to derive the correct quark distribution function in Minkowski space.
In essence the result stems from putting the spectator particle on its complex mass shells. This will be justified by 
careful analytic continuation of Euclidean space amplitudes. Below we consider the Minkowski space calculation of the
generalized parton distribution which cannot be derived by cuts. This will force us to deal with the underlying Wick 
rotation necessary to define the model in Minkowski space.

\begin{figure}
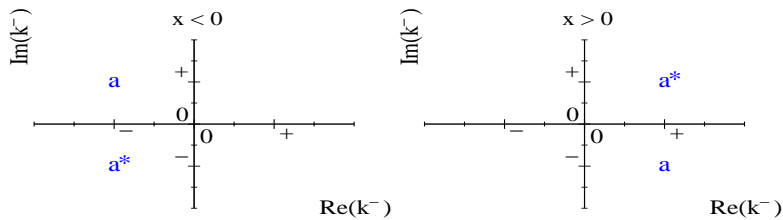

\begin{center}
\epsfig{file=k1.eps,width=2in,height=1.1in}
\epsfig{file=k2.eps,width=2in,height=1.1in}
\caption{Complex light-cone energy plane for the propagator.}
\label{fka}
\end{center}
\end{figure}

In section \ref{euclidean},  the model double distributions were calculated in Euclidean space. Thus to calculate related amplitudes
in Minkowski space, we must Wick rotate in the complex energy plane: $k^4 \to i k^0$. The analytic continuation can be 
viewed alternately in the complex light-cone energy plane. For $k^3 = 0$, the rotation is from the $\Im(k^-)$ axis to the $\Re(k^-)$ axis. 
In the general case, such a correspondence
can only be made precise by considering the Wick rotation in terms of $k^0$ and then boosting to the infinite momentum frame. 
This requires tedious algebraic manipulations and a proliferation of energy poles and time-ordered diagrams. Indeed it is 
easier just to imagine the rotation simply and deal with the light-cone singularities. This is the approach we present. 

Before tackling generalized parton distributions in Minkowski space, let us imagine a simpler fictitious example. Consider 
some well-defined Euclidean space amplitude having only poles at $k^-_a$ and $k^-_{a^*}$, see Eq.~\eqref{apole}. To calculate the
amplitude in Minkowski space, we naively integrate along the $\Re(k^-)$ axis. In general the correct path on which to integrate is 
one which nears the $\Re(k^-)$ axis except for detours around energy poles in the first and third quadrants. Such a path is 
correct since it can be continuously deformed into the Euclidean path. The difference between the naive integration and 
the correct path is a sum of residues of the Wick poles. The energy poles of our fictitious amplitude 
are depicted in Figure \ref{fka}. Their location depends upon the sign of $x = k^+ / P^+$. Thus for this amplitude 
the correct continuation from Euclidean space is
\begin{equation} \label{cont}
\int d\Im (k^-) \longrightarrow \int d \Re (k^-)  +  2 \pi i \big[ \theta(-x) - \theta(x) \big]  \res(k^-_{a^*}).  
\end{equation}
Notice only $k^-_{a^*}$ is a Wick pole; this is expected because we know the limit $\eps \to 0$ can be analytically continued in the naive 
fashion. Closing the contour in the upper-half plane to perform the Minkowski space energy integral\footnote{One must be careful of zero 
modes \cite{Yan:qg} for which $k^+ = 0$. In such cases, the pole lies on the contour at infinity and the integration cannot be performed by
residues. Since our fictitious example is only schematic, we are neglecting the issue of zero modes and are hence excluding amplitudes 
of the form
\begin{equation} \notag
\sum_{\eps = \pm} \int \frac{d^4 k}{(k^2 - m^2 + i \eps)^n}, \quad n>2,
\end{equation} 
which must be handled separately. The example
\begin{equation} \notag
\int \frac{d^4 k}{(k^2 - m^2 + i \eps)^2 (k^2 - m^2 - i \eps)^2}
\end{equation}
is devoid of zero-mode complications and more closely parallels the expressions encountered for GPDs.}
and evaluate our fictitious 
amplitude, we pick up $2 \pi i [\theta(-x) \res (k^-_a) + \theta(x) \res(k^-_{a^*})]$. The net result according to Eq.~\eqref{cont}
is thus
\begin{equation} \notag
2 \pi i \theta(-x) \big[ \res(k_a^-) + \res(k^-_{a^*}) \big].
\end{equation}
Looking back at Figure \ref{fka}, the net result after Wick rotation amounts to both poles lying in the same half-plane; 
or equivalently, we have effectively integrated in either the right- or left-half plane. 

The space-like amplitude\footnote{There are additional complications for time-like amplitudes and for amplitudes involving
unstable bound states. In these cases Wick poles are present even when standard perturbative propagators are used. 
The analysis above must be more carefully considered in these cases where threshold effects are already inherent 
in the analytic continuation to, or from, Euclidean space.} for the generalized parton distribution can now
be continued to Minkowski space and hence be evaluated by projecting onto the light cone. To do so, we refer 
to Figure \ref{ftwist} and insert the non-local light-cone operator $\phi(0) i \overset{\leftrightarrow}{D}{}^+ \phi(z^-)$ in 
place of the local twist-two operators denoted by a cross in the Figure. Here the plus-component picks out the leading-twist contribution 
according to light-cone power counting. Thus in momentum space, we arrive at
\begin{equation} \label{minkGPD}
H(x, \zeta,t) = \frac{2 N / \pi^2}{1 - \zeta / 2} \int d^4 k \; \delta(k^+ - x P^+) \;(2 k^+ + \Delta^+)\; S(k) S(k-P) S(k+\Delta).
\end{equation}
Above we have included the $\zeta$-dependent pre-factor to normalize the action of $\overset{\leftrightarrow}{D}{}^+$ between non-diagonal
states. The overall normalization is then the same as in Eq.~\eqref{bob}. By writing Eq.~\eqref{minkGPD} in Minkowski space, we must also
keep in mind the Wick residues implicitly necessary
so that Eq.~\eqref{minkGPD} is meaningful. In addition to the poles $k^-_a$, $k^-_b$ (given in Eqs.~\eqref{apole} and \eqref{bpole}, 
respectively) and their complex conjugates, the integrand of Eq.~\eqref{minkGPD} also has the poles
\begin{equation} \label{cpole}
k_c^- = - \Delta^- + \frac{(\kperp + \dperp)^2 + m^2}{2 (k^+ + \Delta^+)} - \frac{i \eps}{2 (k^+ + \Delta^+)} 
\end{equation}
and $k^-_{c^*} = (k^-_c)^*$. Carrying out the light-cone energy integration in Eq.~\eqref{minkGPD} as well as adding relevant
residues resulting from the Wick rotation produces (the subtle details of this calculation appear in the appendix)
\begin{equation} \label{GPDres}
H(x, \zeta,t) = - 2 \pi i  \; \theta(x) \theta(\zeta - x) \Big[ \res(k^-_a) + \res(k^-_{a^*}) \Big] 
+ 2 \pi i \; \theta(x - \zeta) \theta(1 - x) \Big[ \res(k^-_b) + \res(k^-_{b^*}) \Big],
\end{equation}
where the residue is of the integrand in Eq.~\eqref{minkGPD}. As a result of the effective relocation of poles
to the same half-plane as their complex conjugates, the resulting GPD Eq.~\eqref{GPDres} is real and vanishes outside $x$ from 
zero to one. 

Using the Weinberg propagator Eq.~\eqref{weinprop}, the residues can be compactly written in terms of relative momenta.
Defining the relative momentum of the final state as
\begin{equation}
\xp = \frac{x - \zeta}{1 - \zeta},  \quad \quad \kpperp = \kperp + (1 - \xp) \dperp,
\end{equation}
and the relative momentum of the photon as
\begin{equation}
\xpp = \frac{x}{\zeta}, \quad \quad \kppperp = \kperp + \xpp \dperp ,
\end{equation}
the light-cone GPD can be expressed in the form
\begin{equation} \label{GPDresult}
( 1 - \zeta / 2) \; H(x,\zeta,t) =  \theta(x) \theta(\zeta - x) \; H_1(x,\zeta,t) + \theta( x - \zeta) \theta(1-x) \; H_2(x,\zeta,t),
\end{equation}
where we have made the abbreviations
\begin{align}
H_1(x,\zeta,t) & = ( 2 \xpp - 1) \frac{N}{2 \pi} \sum_{\eps,\epsp,\epspp = \pm} \int \frac{d\kperp}{\xpp (1 - \xpp) (1 - x)} 
\dw(x,\kperp,\eps,\epspp|M^2) \dw(\xpp,\kppperp, \eps, \epsp |\;t\;) \\
H_2(x,\zeta,t) & = ( 2 x - \zeta) \frac{N}{2 \pi} \sum_{\eps,\epsp,\epspp = \pm} \int \frac{d\kperp}{x (1-x) \xp} 
\dw(x,\kperp, \eps,\epspp |M^2) \dw(\xp,\kpperp,\epsp, \epspp |M^2).
\end{align}

Firstly one can see analytically that the correct quark distribution results at zero momentum transfer, namely
$H_2(x,0,0) = q(x)$, where $q(x)$ is given by Eq.~\eqref{LCqx}. As remarked in section \ref{cutting}, this function is
identical to that obtained from the double distribution Eq.~\eqref{qofx}. Secondly, the resulting light-cone GPD Eq.~\eqref{GPDresult}
agrees numerically with that found from the double distribution, via Eq.~\eqref{JiGPD}, which is plotted in Figure $4$. Finally 
the electromagnetic form factor found from the sum rule
\begin{equation} \label{GPDsum}
F(t) = \int_0^1 dx \; H(x, \zeta, t)
\end{equation}   
agrees numerically with the result of Eq.~\eqref{Fsum} (which is plotted in Figure $3$). The $\zeta$-independence of Eq.~\eqref{GPDsum} 
stems from Lorentz invariance which is present, however, not manifest in Eq.~\eqref{GPDresult}. Thus with the calculation of 
Eq.~\eqref{GPDresult} from analytically continuing to Minkowski space, space-like amplitudes now agree with those calculated 
from the Euclidean space double distribution.

\section{Summary} \label{summy}
Above we consider calculation of amplitudes for space-like processes using a scalar propagator with one pair of complex conjugate 
poles. Although we use a scalar model, generalization to higher spins is clear since the energy denominators are universal.
Moreover the analysis can be extended easily to the case where vertex functions have complex conjugate singularities. 
Such models cannot be directly employed in Minkowski space, they must be analytically continued from Euclidean space. 

In section \ref{minkowski}, the problems of using a propagator with complex conjugate poles in Minkowski space 
are discussed at the level of the quark distribution function. If the model is defined in Minkowski space, one 
will generally violate the support and positivity properties of the quark distribution. Next in section \ref{euclidean}, we
show the model is perfectly well defined in Euclidean space by calculating non-diagonal matrix elements of twist-two operators. 
This leads us to the model's double distribution which we used to calculate parton and generalized parton distributions
as well as the electromagnetic form factor. 

In the remainder of the paper, we investigate how to calculate amplitudes properly in Minkowski space. Firstly amplitudes 
dependent on the imaginary part of some set of diagrams can be calculated by using a straightforward generalization of the
cutting rules. We apply this to calculate the quark distribution function from the handbag diagram 
in the Bjorken limit in section \ref{cutting}. Lastly we consider the analytic continuation of space-like amplitudes from 
Euclidean space. This is complicated by the presence of Wick poles and requires their residues to be appropriately added
when amplitudes are calculated. The details of the GPD calculation appear in the Appendix. 
Resulting functions calculated in Minkowski space after the Wick rotation agree with those
obtained from the model defined in Euclidean space.

This work lays the foundation for calculating light-cone dominated amplitudes using meromorphic model propagators and vertices
constrained by lattice data\footnote{One must proceed with caution: the lattice calculations employ Landau gauge, while we
have tacitly used light-cone gauge above.} and Ward-Takahashi identities.  
Distribution functions for such models could be calculated rigorously since the pole structure of the propagator 
is known and relevant integrals converge in the complex light-cone energy plane. 
As far as light-cone phenomenology is concerned,  resulting expressions would be truly Poincar\'e covariant (as opposed to 
diagonal with respect to the non-interacting operators) and would satisfy 
field-theoretic identities. Filling these two gaps is essential for adequate 
hadronic phenomenology for processes at large momentum transfer.  Moreover, such models would help light-cone methods 
and Dyson-Schwinger studies reach complementary standing.

\begin{acknowledgments}
This work was funded by the U.~S.~Department of Energy, grant: DE-FG$03-97$ER$41014$.  
\end{acknowledgments}

\appendix*

\section{Calculation of the GPD}
Below we derive Eq.~\eqref{GPDres} for the GPD in Minkowski space. The details have been relegated here since there is some subtlety.
In order to evaluate Eq.~\eqref{minkGPD}, which implicitly needs analytic continuation, we must shift the energy integration variable
and define a prescription for dealing with vanishing real parts.  Let us see how these difficulties arise.

In considering the Wick rotation, one is usually only concerned with the single denominator that results from combining 
propagators via Feynman parameters. For the moment, let us ignore the complication of complex conjugate pairs of poles. 
In this case, combining the denominators of Eq.~\eqref{bob} using Feynman parameters results in $[l^2 - D_o(x,y;t) + i \eps]^{-3}$, 
where $D_o$ is given by Eq.~\eqref{dzero}. Since the bound state is stable and $t$ is space-like, $D_o(x,y;t)$ is always positive
and hence there are no Wick poles. Moreover, we have shifted the variable $k^\mu$ to arrive at $l^\mu$.

In analytically continuing the expression with uncombined propagators (and again no complex conjugate poles), we are confronted 
with a problem. The pole $k_b^-$, for example, is shifted by the energy $P^-$. Thus the location of the singularity in the complex plane
will be shifted parallel to the real axis depending upon the relative magnitude of the spectator's kinetic energy and $P^-$. 
In the schematic
example shown in Section \ref{analytic}, the pole $k_a^-$ does not have such a shift. Thus for the $b$ and $c$ poles, the location
of the singularities depicted in Figure \ref{fka} will shift along the real axis (depending on $P^-$ and $\Delta^-$) and there will 
be threshold Wick poles [the threshold is defined when $\Re(k^-_{b,c} = 0$]. This is unphysical: 
we just demonstrated the Wick rotation can be done for the combined denominators without
crossing any poles. To perform the same Wick rotation at the level of separate propagators, we must use the freedom to shift the energy 
variable as well as the stability of the bound state. 

\begin{figure}
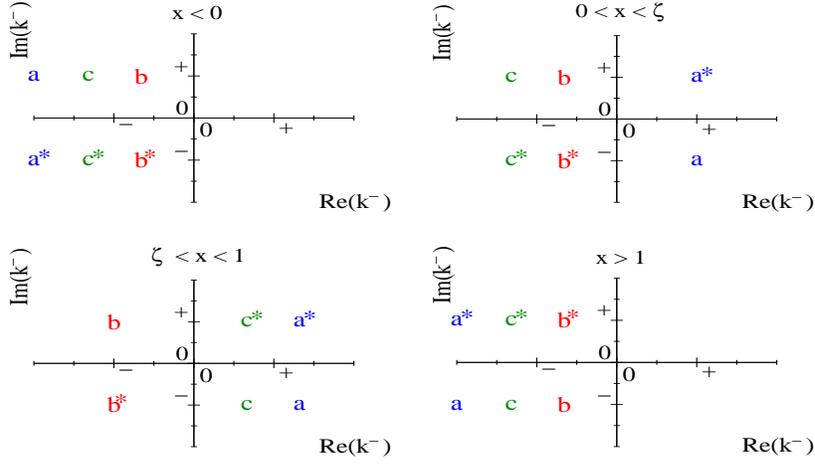

\begin{center}
\epsfig{file=poles1.eps,width=2in,height=1.1in} $\quad$
\epsfig{file=poles2.eps,width=2in,height=1.1in}

\smallskip\smallskip\smallskip\smallskip
\epsfig{file=poles3.eps,width=2in,height=1.1in} $\quad$
\epsfig{file=poles4.eps,width=2in,height=1.1in}
\caption{Complex light-cone energy plane for the shifted poles Eq.~\eqref{newpoles} for the generalized parton distribution.}
\label{fpoles}
\end{center}
\end{figure}
 
On the light-cone, the bound state stability condition can be expressed as
\begin{equation} \label{bssc}
P^- < \frac{\kperp^2_{\text{rel}} + m^2}{2 P^+ x (1 -x)},
\end{equation}
where $\kperp_{\text{rel}} = \kperp -x \mathbf{P}^\perp$ is the relative transverse momentum of the two constituents. 
Because we consider the 
elastic electromagnetic form factor, there is an analogous relation for the final state $P^\prime$. 
Since each propagator contains the kinetic energy of a single particle, the bound state stability condition can never
be utilized without shifting $k^-$. Yet in order to perform such a shift, the real part of one pole must be zero and hence we must
invent a prescription for moving this pole off the Euclidean contour. 

We choose the translation: $k^- \to k^- + \delta k^- + \eta$, where $\eta$ is a positive infinitesimal and
\begin{equation} \label{dk}
\delta k^- = P^- + \frac{\kperp^2 + m^2}{2 P^+ (x - 1)}. 
\end{equation}
The resulting poles of the integrand in Eq.~\eqref{minkGPD} we denote
\begin{equation} \label{newpoles}
\begin{cases}
\tilde{k}^-_a = k^-_a - \delta k^-  = - P^- + \frac{\kperp^2 + m^2}{2 P^+ x (1-x)} - \frac{i \eps}{2 P^+ x}\\
\tilde{k}^-_b = k^-_b - \delta k^- - \eta = - \eta - \frac{i \eps}{2 P^+ (x - 1)} \\
\tilde{k}^-_c = k^-_c - \delta k^-  = - P^{\prime-} + \frac{\kpperp^2 + m^2}{2 P^{\prime+} \xp (1- \xp) } - 
\frac{i \eps}{2 P^+ (x - \zeta)}
\end{cases},
\end{equation}
and similarly for their complex conjugate partners. Notice imaginary parts of the poles are unaffected by the energy translation. 
The $\eta$ prescription has displaced the resulting spectator pole away from the Euclidean path independent of $x$. Moreover the
$a$ and $c$ poles have non-zero real parts; so $\eta$ has been set to zero for these poles above.

Using the expressions for the new poles Eq.~\eqref{newpoles} and the bound state stability condition Eq.~\eqref{bssc}, we can 
determine the quadrant location of the singularities independent of $P^-$ and $\Delta^-$ (see footnote $4$). These quadrant locations 
are depicted for the full range of $x$ in Figure \ref{fpoles}. Accordingly poles in the first and third quadrants are Wick poles. As we saw in Section
\ref{analytic}, the net result
of the analytic continuation is to evaluate the integral by effectively closing the contour in the right- or left-half plane.  
Hence the GPD Eq.~\eqref{minkGPD} is
\begin{equation} \label{GPDrestilde}
H(x, \zeta,t) = - 2 \pi i  \; \theta(x) \theta(\zeta - x) \Big[ \res(\tilde{k}^-_a) + \res(\tilde{k}^-_{a^*}) \Big] 
+ 2 \pi i \; \theta(x - \zeta) \theta(1 - x) \Big[ \res(\tilde{k}^-_b) + \res(\tilde{k}^-_{b^*}) \Big].
\end{equation}
Evaluating the residues in Eq.~\eqref{GPDrestilde} yields the result of Section \ref{analytic}, namely  Eq.~\eqref{GPDresult} which 
is algebraically equivalent to Eq.~\eqref{GPDres}.  

Notice from Figure \ref{fpoles}, other $\eta$ prescriptions for the shift, such as $+\frac{\eta}{2 P^+ (x - 1)}$, lead to an incorrect
result for the case when there are no complex conjugate pairs. The figure shows that the infinitesimal prescription must be positive 
and independent of the sign of $x$, $x - \zeta$, \emph{etc}, in order to reproduce the familiar result. It is interesting to note
that for $x > 1$ in the case without conjugate pairs, all poles of the integrand are Wick poles. Though since the integral is convergent,
the sum of these Wick residues vanishes.

The interested reader can verify that the alternate shifts which use the same pole prescription
\begin{equation}
\begin{cases}
	k^- \to k^- + \frac{\kperp^2 + m^2}{2 P^+ x} + \eta \\
	k^- \to k^- - \Delta^- + \frac{(\kperp+ \dperp)^2 + m^2}{2 P^+ ( x - \zeta)} + \eta
\end{cases}
\end{equation}
also yield the correct results provided $t$ is space-like.


\begin{thebibliography}{99}
\bibitem{Lepage:1980fj}
G.~P.~Lepage and S.~J.~Brodsky,
Phys.\ Rev.\ D {\bf 22}, 2157 (1980).

\bibitem{Muller:1998fv}
D.~M\"uller, D.~Robaschik, B.~Geyer, F.~M.~Dittes and J.~Ho\v{r}ej\v{s}i,
Fortsch.\ Phys.\  {\bf 42}, 101 (1994);

X.-D.~Ji,
Phys.\ Rev.\ Lett.\  {\bf 78}, 610 (1997);
Phys.\ Rev.\ D {\bf 55}, 7114 (1997);

A.~V.~Radyushkin,
Phys.\ Lett.\ B {\bf 380}, 417 (1996);
{\bf 385}, 333 (1996).

\bibitem{Ji:1998pc}
X.-D.~Ji,
J.\ Phys.\ G {\bf 24}, 1181 (1998);

A.~V.~Radyushkin,
hep-ph/0101225;

K.~Goeke, M.~V.~Polyakov and M.~Vanderhaeghen,
Prog.\ Part.\ Nucl.\ Phys.\  {\bf 47}, 401 (2001);

A.~V.~Belitsky, D.~M\"uller and A.~Kirchner,
Nucl.\ Phys.\ B {\bf 629}, 323 (2002).


\bibitem{Brodsky:1997de}
S.~J.~Brodsky, H.~C.~Pauli and S.~S.~Pinsky,
Phys.\ Rept.\  {\bf 301}, 299 (1998).

\bibitem{Dalley:1998bj}
S.~Dalley and B.~van de Sande,
Phys.\ Rev.\ D {\bf 59}, 065008 (1999);
hep-ph/0212086;

M.~Burkardt and S.~K.~Seal,
Phys.\ Rev.\ D {\bf 65}, 034501 (2002).

\bibitem{Ji:2002xn}
X.-D.~Ji, J.-P.~Ma and F.~Yuan,
Nucl.\ Phys.\ B {\bf 652}, 383 (2003);
hep-ph/0301141;
hep-ph/0304107.


\bibitem{Maris:2003vk}
P.~Maris and C.~D.~Roberts,
nucl-th/0301049.

\bibitem{Atkinson:1978tk}
D.~Atkinson and D.~W.~Blatt,
Nucl.\ Phys.\ B {\bf 151}, 342 (1979);

C.~J.~Burden, C.~D.~Roberts and A.~G.~Williams,
Phys.\ Lett.\ B {\bf 285}, 347 (1992);

G.~Krein, C.~D.~Roberts and A.~G.~Williams,
Int.\ J.\ Mod.\ Phys.\ A {\bf 7}, 5607 (1992);

U.~Habel, R.~Konning, H.~G.~Reusch, M.~Stingl and S.~Wigard,
Z.\ Phys.\ A {\bf 336}, 423 (1990);

M.~Stingl,
Z.\ Phys.\ A {\bf 353}, 423 (1996);

P.~Maris,
Phys.\ Rev.\ D {\bf 52}, 6087 (1995);

V.~N.~Gribov,
Eur.\ Phys.\ J.\ C {\bf 10}, 91 (1999).

\bibitem{Bhagwat:2003wu}
M.~S.~Bhagwat, M.~A.~Pichowsky and P.~C.~Tandy,
Phys.\ Rev.\ D {\bf 67}, 054019 (2003).

\bibitem{Bhagwat:2003vw}
M.~S.~Bhagwat, M.~A.~Pichowsky, C.~D.~Roberts and P.~C.~Tandy,
nucl-th/0304003.

\bibitem{unp}
R.~Alkofer, W.~Detmold, C.~Fisher and P.~Maris, in preparation.

\bibitem{Kusaka:1996vm}
K.~Kusaka, G.~Piller, A.~W.~Thomas and A.~G.~Williams,
Phys.\ Rev.\ D {\bf 55}, 5299 (1997).

\bibitem{Hecht:2000xa}
M.~B.~Hecht, C.~D.~Roberts and S.~M.~Schmidt,
Phys.\ Rev.\ C {\bf 63}, 025213 (2001).

\bibitem{Radyushkin:1997ki}
A.~V.~Radyushkin,
Phys.\ Rev.\ D {\bf 56}, 5524 (1997);
{\bf 59}, 014030 (1999).

\bibitem{Mukherjee:2002gb}
A.~Mukherjee, I.~V.~Musatov, H.~C.~Pauli and A.~V.~Radyushkin,
Phys.\ Rev.\ D {\bf 67}, 073014 (2003).

\bibitem{Tiburzi:2002kr}
B.~C.~Tiburzi and G.~A.~Miller,
Phys.\ Rev.\ D {\bf 67}, 013010 (2003).

\bibitem{Golec-Biernat:1998ja}
K.~J.~Golec-Biernat and A.~D.~Martin,
Phys.\ Rev.\ D {\bf 59}, 014029 (1999).

\bibitem{Teryaev:2001qm}
O.~V.~Teryaev,
Phys.\ Lett.\ B {\bf 510}, 125 (2001).

\bibitem{Polyakov:1999gs}
M.~V.~Polyakov and C.~Weiss,
Phys.\ Rev.\ D {\bf 60}, 114017 (1999).

\bibitem{Belitsky:2000vk}
A.~V.~Belitsky, D.~M\"uller, A.~Kirchner and A.~Sch\"afer,
Phys.\ Rev.\ D {\bf 64}, 116002 (2001).

\bibitem{Tiburzi:2002tq}
B.~C.~Tiburzi and G.~A.~Miller,
hep-ph/0212238.

\bibitem{Mankiewicz:1997uy}
L.~Mankiewicz, G.~Piller and T.~Weigl,
Eur.\ Phys.\ J.\ C {\bf 5}, 119 (1998).

\bibitem{Tiburzi:2001je}
B.~C.~Tiburzi and G.~A.~Miller,
Phys.\ Rev.\ D {\bf 65}, 074009 (2002).

\bibitem{Yan:qg}
T.-M.~Yan,
Phys.\ Rev.\ D {\bf 7}, 1780 (1973).

\end{thebibliography}
\end{document}